\documentclass[a4paper,aps,pre,twocolumn,groupedaddress,showkeys,showpacs,floats,floatats,floatfix]{revtex4-1}
\usepackage[utf8]{inputenc}
\usepackage[english]{babel}
\usepackage{graphicx}
\usepackage{dcolumn} 
\usepackage{bm}
\usepackage{natbib}
\usepackage{latexsym}
\usepackage{mathrsfs}
\usepackage{amssymb}
\usepackage{amsmath}
\usepackage{amscd}
\usepackage{color}
\usepackage{pifont}
\usepackage{pstricks,pst-node,pst-text,pst-3d}
\usepackage{verbatim}
\usepackage{ulem}
\usepackage[T1]{fontenc}
\usepackage{hyperref}
\hypersetup{
  colorlinks   = true, 
  urlcolor     = black, 
  linkcolor    = red, 
  citecolor    = blue 
}
\definecolor{aquamarine}{rgb}{0.5, 1.0, 0.83}
\definecolor{blue-violet}{rgb}{0.54, 0.17, 0.89}

\newrgbcolor{Red}{1.0 0.0 1.0}
\begin{document}
\title{Optimal ratchet current for elastically interacting particles}
\author{Rafael M.~da Silva$^{1,2}$}
\email{rmarques@fisica.ufpr.br}

\author{Cesar Manchein$^3$}
\email{cesar.manchein@udesc.br}
\author{Marcus W.~Beims$^{2}$}
\email{mbeims@fisica.ufpr.br}
\affiliation{$^1$Departamento de Física, Universidade Federal da
  Paraíba, 58051-900 João Pessoa, PB, Brazil}
\affiliation{$^2$Departamento de F\'\i sica, Universidade Federal do
  Paran\'a, 81531-980 Curitiba, PR, Brazil}
\affiliation{$^3$Departamento de F\'\i sica, Universidade do Estado de
  Santa Catarina, 89219-710 Joinville, SC, Brazil} 
\date{\today}
%
\begin{abstract}
In this work we show that optimal ratchet currents of two interacting 
particles are obtained when stable periodic motion is present. By
increasing the coupling strength between identical ratchet maps, it is
possible to find, for some parametric combinations, current reversals,
hyperchaos, multistability, and duplication of the periodic motion in
the parameter space. Besides that, setting a fixed value for the
current of one ratchet it is possible to induce a positive/negative/null
current for the whole system in certain domains of the  parameter space.
\end{abstract}
%
\maketitle

{\bf Ratchet models are well-known prominent candidates to describe
  the transport phenomenon in nature in the absence of external
  bias. The ratchet effect is defined as the rectification of an
  external net-zero force to obtain a directional transport of
  particles in spatially periodic media. While single ratchet systems
  have been extensively studied in the literature, the dynamics in
  coupled ratchets is almost unknown. In this work we exhaustively
  explore the parameter space of two elastically coupled ratchet model
  and characterize the  global dynamics of the whole system for a
  symmetric coupling configuration. The parameters of the system are
  the coupling strength of elastic interaction between the ratchets,
  the damping factor, and the ratchet potential amplitude. Such simple
  configuration is enough to produce competing influences between the
  ratchets and leads to the appearance of complex behaviors as
  hyperchaotic dynamics, current reversals and multistability observed
  in phase and parameter spaces.} 

\section{Introduction}
\label{intro}
Ratchets are physical devices constructed in such a clever way that 
they are able to transport particles with nonzero macroscopic  
velocity in the absence of macroscopic external bias. They only
operate with success when time and spatial symmetries are
simultaneously broken. The motion of single Brownian particles in
ratchet-like devices has attracted great attention in the last 
decades due to its great ability to describe systems and phenomena as 
molecular motors \cite{PhysRevLett.77.194, REIMANN200257,Astumian917},
coupled Josephson junctions \cite{Zapata96}, mass separation
\cite{Kettner00} and transport of solitons in crystals
\cite{Brox17}, among many others. In the context of molecular motors 
(protein molecules that perform essential tasks to the life of the
organism like muscle contraction, intracellular transport, and cell
division), the ratchet model has helped to understand how they operate
\cite{AstumianHanggi}. A crucial advance in the understanding of how
optimal ratchets currents are generated was only obtained when the
relation between optimal ratchet currents (RCs) and the Isoperiodic
Stable Structures (ISSs) was established \cite{Alan-2011}. The ISSs,
observed in many dissipative continuous- and discrete-time dynamical
systems   
\cite{Fraser82,Mira91,Gallas93,Bonatto05,Bonatto07,Stoop10,DACOSTA20161610},
delimit the range of available parameters which lead to regular
orbits. Most importantly, the ISSs appear along preferred directions in 
the parameter space furnishing a guide to follow the optimal ratchet 
current \cite{Alan-2011}. 

The relevance of parameters chosen inside ISSs for the generation of
optimal currents in {\it coupled} ratchet devices is completely
unknown. The main goal of the present work is to characterize this 
relevance. Realistic systems are composed of many interacting
particles and it is not clear what is the contribution of the mutual
particle interaction on the ratchet current. Thus, in this work we 
start showing the effects of the elastic interaction between two
ratchet devices on the current inside ISSs. Some transport
properties of an elastically coupled lattice of  particles in a
periodically flashing ratchet potential has already been examined  
\cite{Chen05}. They demonstrated that mutual couplings among particles
may improve the transport efficiency when the coupling strength
overcomes a threshold and the interaction between the particles
clearly influences the directed motion. An effective potential for the
center-of-mass of particles has been proposed \cite{WANG200413} in
order to understand this behaviour. 

This paper is organized as follows. Section~\ref{mod} presents the
model of coupled maps, summarizes some properties of the uncoupled
ratchet map and defines the ratchet current. Discussions and results
are given in Sec.~\ref{eql}  for two identical coupled ratchet 
maps, and in Sec.~\ref{diff} for distinct coupled ratchet maps. 
Finally in Sec.~\ref{conclusion} we present the conclusions.

\section{Two coupled-ratchets model}
\label{mod}
The model studied in this work is composed of two-coupled Ratchet Maps
(RMs) \cite{Wang-2007,Alan-2011} which may present distinct dynamical
properties. The coupling force between the ratchets is elastic
\cite{Kurths-2010,Levien-2015} according to:
\renewcommand{\arraystretch}{1.5}
\begin{equation}
\label{rm-el}
\mbox{\bf RM}^{(1)}\left\{
\begin{array}{llll}
p^{\prime}_{1} =& \gamma_{1} p_{1} + K_{1} [\sin (x_{1}) + a \hspace{0.05cm} 
\sin (2 x_{1} + \phi)]\\ &+\, c\,(x_{2} - x_{1}), \\
x^{\prime}_{1} =& x_{1} + p^{\prime}_{1}, \\
\end{array}
\right.
\end{equation}
\begin{equation}
\mbox{\bf RM}^{(2)}\left\{
\begin{array}{llll}
p^{\prime}_{2} =& \gamma_{2} p_{2} + K_{2} [\sin (x_{2}) + a \hspace{0.05cm} 
\sin (2 x_{2} + \phi)]\\ & +\, c\,(x_{1} - x_{2}), \\
x^{\prime}_{2} = &x_{2} + p^{\prime}_{2}, \nonumber \\
\end{array}
\right.
\end{equation}
%
%
%
\noindent where $K_{i}$, for $i=1,2$, are the nonlinearity parameters and
$c$ is the coupling constant. The prime represents the discrete time
evolution. This model describes the dynamics of two coupled particles
that move on the $x$ direction with $x\in (-\infty,+\infty)$ in a
asymmetric potential, while $p$ is the conjugate momentum of $x$ and  
$\gamma_i \in [0,1]$ represents the dissipation of particle $i$. The
spatial symmetry is broken when $a \neq 0$ and $\phi \neq l \pi$,
where $l$ is an integer. In the present work we kept fixed the
parameters $a=0.5$ and $\phi=\pi/2$ \cite{Alan-2011,RMS-2018454} and
studied the two dimensional parameter space $(K_{2},\gamma_{2})$ for
the map $\mbox{\bf RM}^{(2)}$ using different parametric combinations 
$(K_{1},\gamma_{1})$ for the map $\mbox{\bf RM}^{(1)}$ setting different 
intensities for $c$. In Section~\ref{eql} the case $(K_{1},\gamma_{1})=
(K_{2},\gamma_{2})$ was analysed and in Section~\ref{diff} the parameter 
combination $(K_{1},\gamma_{1})$ was kept fixed, allowing to choose the 
momentum of the particle $i=1$.

The quantity that gives us insights about transport properties of the
$i$-th RM is the {\it Ratchet Current} (RC) $\mathcal{J}_i$, defined
as a double average of the momentum $p_i$:
\begin{equation}
  \label{current}
  \mathcal{J}_i = \dfrac{1}{M} \displaystyle \sum_{k=1}^M \left[\dfrac{1}{N} 
    \displaystyle \sum_{n=1}^N p_{i}^{(k,n)}  \right], 
\end{equation}
where $M$ is the number of initial conditions (ICs) and $N$ the total 
iteration  time. The ICs must be equally distributed between the interval 
$(x_{i}^{\mbox{\tiny min}},x_i^{\mbox{\tiny max}})=
(p_i^{\mbox{\tiny min}},p_i^{\mbox{\tiny max}})=(-2\pi,2\pi)$,
since they can not establish an preferred direction. For 
the system~(\ref{rm-el}), the {\it total ratchet current} (TRC) 
$\mathcal{J}_T$ is give by  
\begin{equation}
  \label{tcurrent}
  \mathcal{J}_T = \mathcal{J}_1 + \mathcal{J}_2.
\end{equation}
By this definition, the TRC tends to zero when $p_1 = -p_2$, while the 
maximum magnitude for TRC is obtained when $p_1$ and $p_2$ have the same 
direction and assume the maximum value.

\section{Coupling of two identical RMs}
\label{eql}

To determine the influence of the coupling strength $c$ on the
dynamics of the system~(\ref{rm-el}), we start plotting the parameter
space $(K_2,\gamma_2)$ with colors representing the respective TRC
$\mathcal{J}_T$ (see color bars in the figures). The first case
considered, shown in Fig.~\ref{fig1}, consists of a system of two
identical coupled RMs for which $(K_1,\gamma_1)=(K_2,\gamma_2)$. Figure 
\ref{fig1}(a) shows $\mathcal{J}_T$ for the uncoupled case $c=0$. The 
configuration observed in the parameter space is identical to that 
obtained for a single map (see {\it e.g.} Refs.~\cite{Alan-2011, 
RMS-2018454}). The difference is that the colors in Fig.~\ref{fig1} 
represent the TRC for the system~(\ref{rm-el}). In the uncoupled case 
$\mathcal{J}_T=2\mathcal{J}_1=2\mathcal{J}_2$, since the maps are equal. 
As can be seen in the palette positioned to the right of Fig.~\ref{fig1}, 
black color represents TRCs close to zero, increasing positive TRCs are 
found in regions with colors going from cyan to blue, and increasing 
negative TRCs are identified by the colors changing from yellow to red.
\begin{widetext}
$\quad$
\begin{figure}[b]
  \centering
  \includegraphics*[width=1.0\columnwidth]{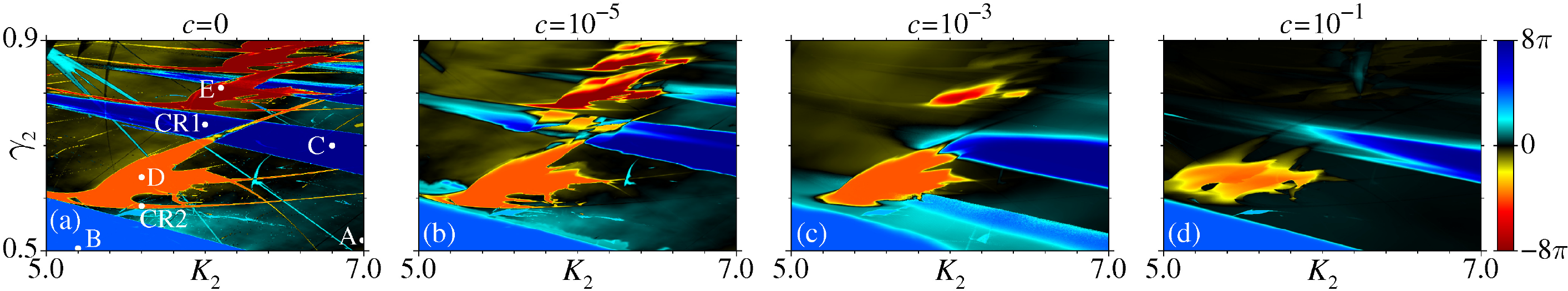}
  \caption{(Color online) Parameter space $(K_2,\gamma_2)$ with colors
    representing the TRC $\mathcal{J}_T$ for the system (\ref{rm-el})
    in (a) ($c=0$), (b) ($c=10^{-5}$), (c) ($c=10^{-3}$), and (d)
    ($c=10^{-1}$), considering the coupling of two identical RMs
    [$(K_1,\gamma_1)=(K_2,\gamma_2)$]. For each parametric combination
    $(K_2,\gamma_2)$, $\mathcal{J}_T$ was obtained using $10^4$ ICs
    and $N=5\times 10^4$ iterations after a transient time of $8\times
    10^4$ iterations.} 
  \label{fig1}
\end{figure}
\end{widetext}
It is well established in the literature that, for the uncoupled case, 
most regions with zero RCs correspond to a chaotic dynamics, while 
parametric combinations that lead to optimized values of RCs are 
delimited by the ISSs~\cite{Alan-2011}. 

Increasing the coupling strength $c$, it is possible to note in 
Figs.~\ref{fig1}(b), \ref{fig1}(c), and \ref{fig1}(d) that the coupling 
tends to destroy the ISSs, starting from their antennas, decreasing the 
amount of parametric combinations that generates non-zero RCs. Such
destruction of the ISSs is similar to what is observed in ratchet
subjected to noise or temperature (for more details see
\cite{Alan-2013,ham17,RMS-2018454}). Besides that, it is possible  
to observe a small translation in parameter space from those ISSs
which resist to larger coupling strengths and contain positive or
negative TRCs.   

\subsection{Current Reversal}
In the system composed of two identical RMs we can observe that,
increasing the coupling intensity $c$, the direction of particle's motion 
changes in some regions of the parameter space. To demonstrate this 
phenomenon (also exhibited in continuous-time dynamical systems of two
interacting ratchets \cite{Kurths-2010, Li2016}), we highlighted two
parametric combinations: ($i$) $(K_1,\gamma_1)=(K_2,\gamma_2)=(6.0,0.74)$, 
indicated in Fig.~\ref{fig1}(a) by the white point CR1; and ($ii$)
$(K_1,\gamma_1)=(K_2,\gamma_2)=(5.6,0.585)$, indicated in Fig.~\ref{fig1}(a) 
by the white point CR2. To analyze the current reversal, we kept fixed these 
parametric combinations and increased slightly the coupling strength $c$. For 
each value of $c$, we performed the calculation of $\mathcal{J}_T$ using 
$10^4$ ICs equally distributed between the interval 
$(x_{i}^{\mbox{\tiny min}},x_i^{\mbox{\tiny max}})=(p_i^{\mbox{\tiny min}},
p_i^{\mbox{\tiny max}})=(-2\pi,2\pi)$ and $N=5\times 10^4$ iterations after a 
transient time of $8\times 10^4$ iterations.  

\begin{figure}[!t]
  \centering
  \includegraphics*[width=0.98\columnwidth]{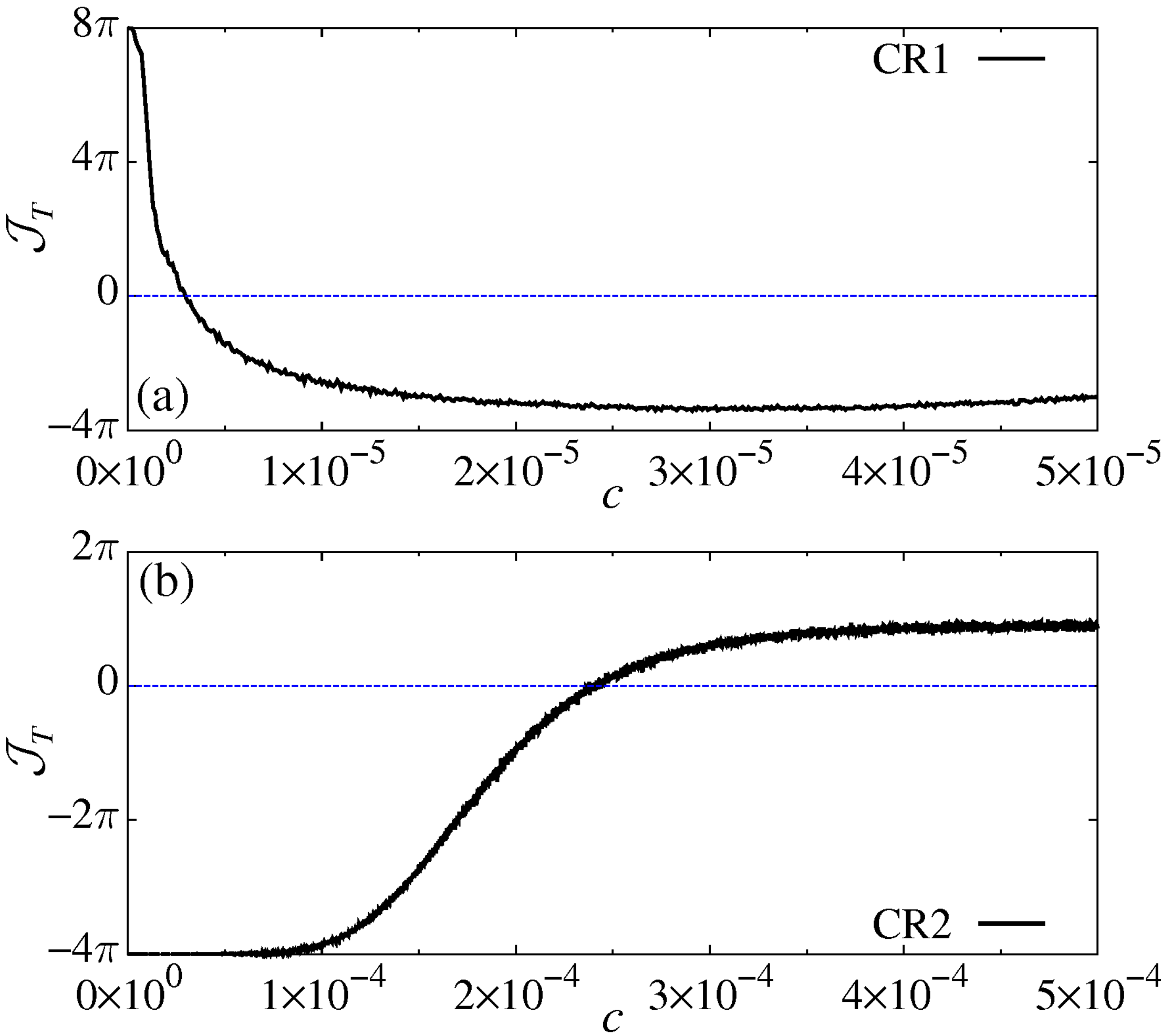}
  \caption{(Color online) TRC $\mathcal{J}_T$ as a function of $c$ for
  the system~(\ref{rm-el}) considering the coupling of two identical RMs 
  with $(K_1,\gamma_1)=(K_2,\gamma_2)=(6.0,0.74)$ in (a) [white point CR1 
  in Fig.~\ref{fig1}(a)] and $(K_1,\gamma_1)=(K_2,\gamma_2)=(5.6,0.585)$ in 
  (b) [white point CR2 in Fig.~\ref{fig1}(a)]. For each value of $c$, 
  $\mathcal{J}_T$ was obtained using $10^4$ ICs and $N=5\times 10^4$ 
  iterations after a transient time of $8\times 10^4$ iterations. } 
  \label{fig2}
\end{figure}

In Fig.~\ref{fig2}(a) the behavior of $\mathcal{J}_T$ as a function of $c$ 
for the case ($i$) is plotted. For this parametric combination, 
$\mathcal{J}_T \approx 8\pi$ for $c=0$, since we have the sum of two uncoupled
particles with RC $\mathcal{J}_i \approx 4\pi$. The current reversal from 
$\mathcal{J}_T>0$ to $\mathcal{J}_T<0$ occurs for $c\approx 3\times 10^{-6}$. 
The negative TRC with maximum magnitude is obtained for $c\approx 2.9\times 
10^{-5}$, and is $\mathcal{J}_T\approx -10.745215$. For this case, increasing 
the value of $c$, the TRC tends to zero. In Fig.~\ref{fig2}(b) the same analysis
is shown but for the case ($ii$), for which we have a transition from 
$\mathcal{J}_T<0$ to $\mathcal{J}_T>0$ that occurs for $c\approx 2.4\times 
10^{-4}$. For this parametric combination, $\mathcal{J}_T \approx -4\pi$ if $c=0$ 
and for $c>2.4\times 10^{-4}$ the value of TRC stabilizes around $+\pi$. 
Additional simulations (not shown here) demonstrate that the TRC remains around 
$+\pi$ until $c \approx 10^{-2}$ and, after this coupling strength, 
$\mathcal{J}_T$ tends to zero. 

\subsection{Duplication of ISSs}

Using strong coupling intensities in the system~(\ref{rm-el}) we
can observe that the domain of the parameter space with negative RCs
is duplicated, as shown in Fig.~\ref{fig1}(d) for $c=10^{-1}$. Recently, 
similar findings were reported in time discrete dynamical systems by using 
external periodic perturbations to generate multiple attractors in phase 
space and ISSs in the parameter space \cite{rafael17-2, rafael17-1}. 
Increasing the intensity of the external perturbation, the regular 
structures start to separate from each other and an effective enlargement 
of the available stable domain in the parameter space is obtained. This 
methodology was successfully applied in the RM to retain thermal effects 
and to increase the area of the parameter space that leads to optimal RCs
\cite{RMS-2018454}. In the case of continuous systems, new stable attractors 
are not created but the steering of the existing multiple attractors in 
phase space was used to increase the regular region of the parameter space 
for the Langevin equation and for the Chua's electronic 
circuit~\cite{PRE98-032210}. This results lead us to conclude that 
duplications of stable regimes are closely related to multistability. 

\begin{figure}[!t]
  \centering
  \includegraphics*[width=1.00\columnwidth]{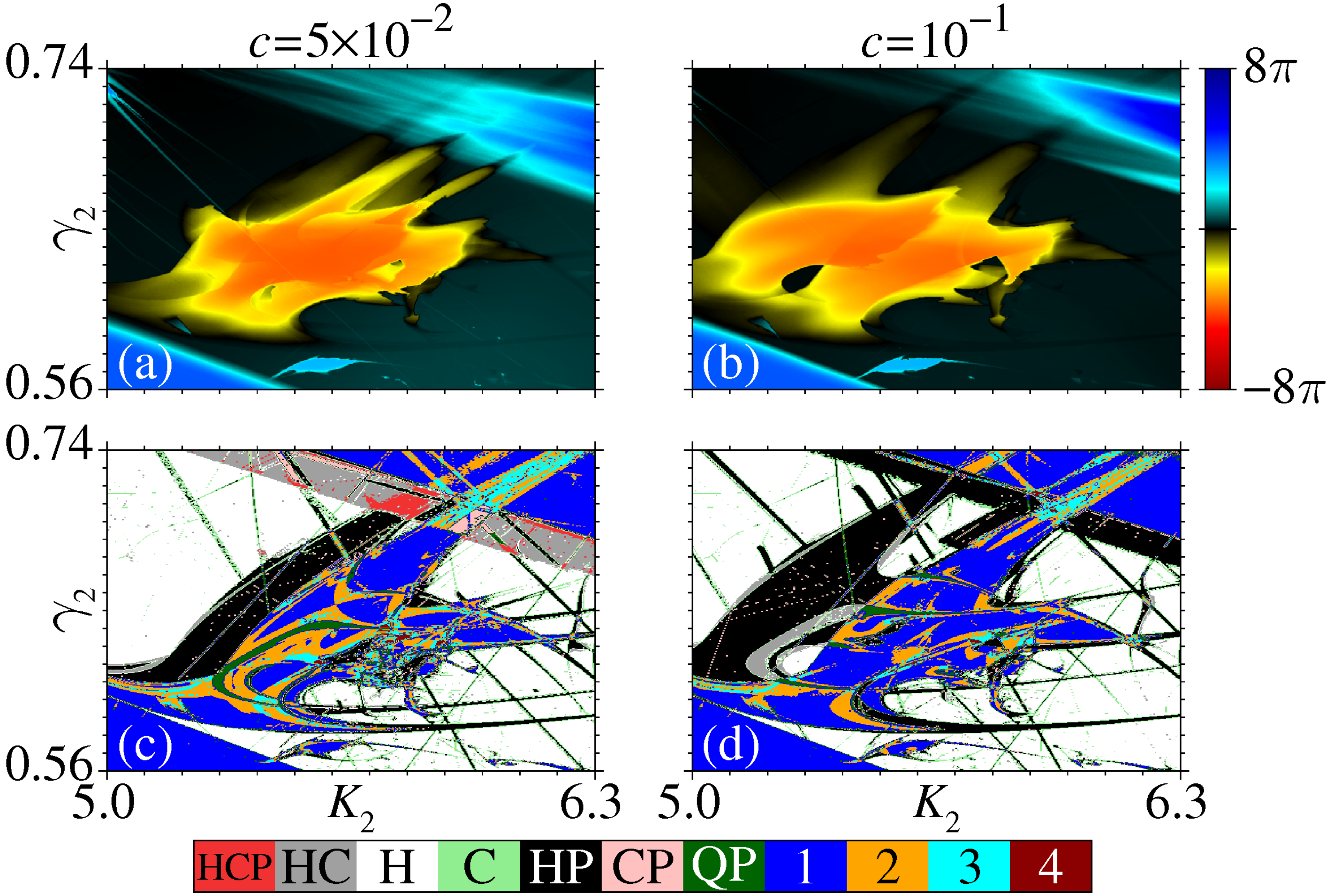}
  \caption{(Color online) Parameter space $(K_2,\gamma_2)$ with colors
    representing the TRC $\mathcal{J}_T$ for the system (\ref{rm-el})
    in (a) ($c=5\times 10^{-2}$) and (b) ($c=10^{-1}$), and the
    different classes of attractors found in each parametric
    combination in (c) ($c=5\times 10^{-2}$) and (d) ($c=10^{-1}$).}   
  \label{fig3}
\end{figure}

To investigate the process behind the duplication in the case of the 
coupled system~(\ref{rm-el}), we focus on the parametric range presented 
in Fig.~\ref{fig3} for couplings $c=5\times 10^{-2}$ and $c=1\times 10^{-1}$. 
We can observe in Figs.~\ref{fig3}(a) and \ref{fig3}(b) (see color bar) that 
increasing the value of $c$ a second parametric domain with negative RC 
(yellow and red) arises and these two identical regions start to move away from 
each other. In Figs.~\ref{fig3}(c) and \ref{fig3}(d) the same range was analyzed, 
but the colors now represent the possible combinations of different
classes of attractors that can be found for the same point
$(K_1,\gamma_1)=(K_2,\gamma_2)$ in the parameter space. To obtain this
information, we use $81$ ICs for each parametric combination and look
at the values of the Lyapunov exponents (LEs) of the resulting
trajectories. For a periodic attractor (P), $\lambda_i<0$ for
$i=1,2,3,4$. A chaotic attractor (C) takes place if $\lambda_1>0$ and  
$\lambda_i<0$, for $i=2,3,4$, while a hyperchaotic attractor (H) is
typically defined as a chaotic attractor with at least two positive
Lyapunov exponents. In addition to these three, there is another type
of dynamical motion that is common: the quasiperiodic motion. A
bounded orbit that is not asymptotically periodic and that does not
exhibit sensitive dependence on initial conditions is called
quasiperiodic \cite{asy96}. Thus, for a quasiperiodic attractor (QP) we
have $\lambda_1=0$ and $\lambda_i<0$ for $i=2,3,4$, or
$\lambda_1=\lambda_2=0$ and $\lambda_i<0$ for $i=3,4$.
Given the above definitions, in Figs.~\ref{fig3}(c) and \ref{fig3}(d)  
it is  possible to find, {\it for the same point in the parameter
  space,} the following classes of attractors: 
\begin{itemize}
\item {\bf HCP:} hyperchaotic, chaotic, and periodic attractors;
\item {\bf HC:} hyperchaotic and chaotic attractors;
\item {\bf H:} only hyperchaotic attractors; 
\item {\bf C:} only chaotic attractors;
\item {\bf HP:} hyperchaotic and periodic attractors;
\item {\bf CP:} chaotic and periodic attractors;
\item {\bf QP:} quasiperiodic and periodic attractors;
\item {\bf 1, 2, 3,} and {\bf 4} different periodic attractors. 
\end{itemize}
\noindent Some of these combinations of coexisting attractors can be 
seen in Fig.~\ref{fig4}, that shows the projection of the phase space 
$(x_1,p_1,x_2,p_2)$ of the system~(\ref{rm-el}) in the plane $(x_2,p_2)$, 
considering the coupling strength $c=5\times 10^{-2}$. 

In Fig.~\ref{fig4}(a) the hyperchaotic (blue), the chaotic (yellow) and the
periodic (black) attractors found for $(K_1,\gamma_1)=(K_2,\gamma_2)=(5.8,0.71)$ 
are plotted. This parametric combination is indicated in Fig.~\ref{fig3}(c) by
the red color (HCP). Figure~\ref{fig4}(b) shows the hyperchaotic (blue) and the 
periodic (black) attractors that can be found for $(K_1,\gamma_1)=(K_2,\gamma_2)
=(5.4,0.65)$, parametric combination indicated in Fig.~\ref{fig3}(c) by the black 
color (HP). A quasiperiodic attractor (red) is plotted in Fig.~\ref{fig4}(c) along 
with a periodic attractor (black), both found for $(K_1,\gamma_1)=(K_2,\gamma_2)=
(5.3,0.61)$. The regions of the parameter space for which we found at
least one quasiperiodic attractor are indicated in Figs.~\ref{fig3}(c)
and \ref{fig3}(d) by the green color (QP). These domains are located
inside the ISSs, next to the  parameters for which period doubling
bifurcations occur for the uncoupled case. The coexistence of two
periodic attractors that occurs for  
\begin{widetext}
$\quad$
\begin{figure}[!b]
  \centering
  \includegraphics*[width=0.97\columnwidth]{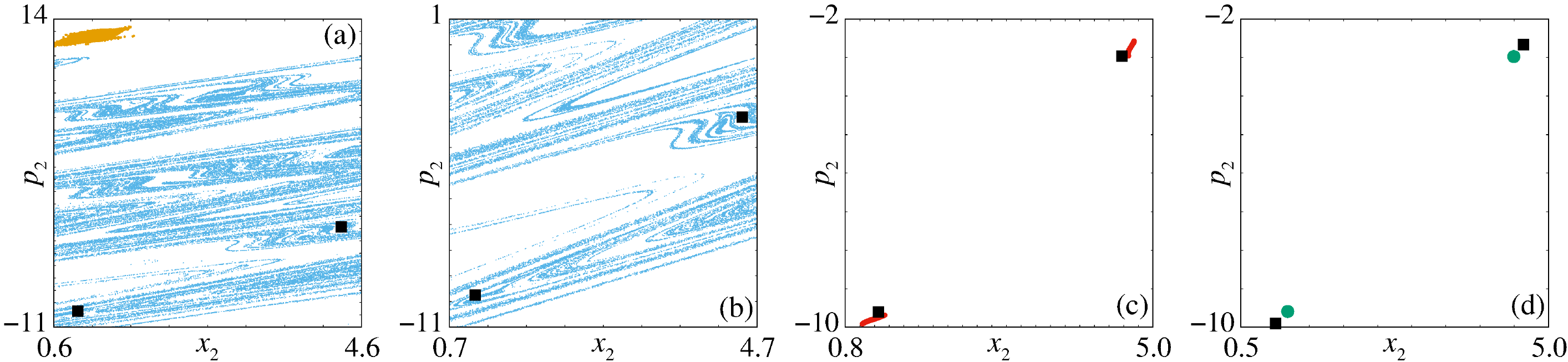}
  \caption{(Color online) Projection of the four-dimensional phase
    space $(x_1,p_1,x_2,p_2)$ in the plane $(x_2,p_2)$ showing the
    different classes of attractors found in (a)
    $(K_1,\gamma_1)=(K_2,\gamma_2)=(5.8,0.71)$: HCP - hyperchaotic (blue), 
    chaotic (yellow) and periodic (black) attractors; (b) $(K_1,\gamma_1)=
    (K_2,\gamma_2)=(5.4,0.65)$: HP - hyperchaotic (blue) and periodic (black) 
    attractors; (c) $(K_1,\gamma_1)=(K_2,\gamma_2)=(5.3,0.61)$: QP - 
    quasiperiodic (red) and periodic (black) attractors; and (d) $(K_1,\gamma_1) 
    =(K_2,\gamma_2)=(5.4,0.63)$: 2 different periodic attractors (black and 
    green), all cases using $c=5\times 10^{-2}$.}
  \label{fig4}
\end{figure}
%
\end{widetext}
$(K_1,\gamma_1)=(K_2,\gamma_2)=(5.4,0.63)$ is shown in
Fig.~\ref{fig4}(d). To identify different periodic attractors, we
compared the largest LE $\lambda_1$ and the total time average
$\langle p_1 \rangle + \langle p_2 \rangle$ obtained for each
attractor between them. These quantities are calculated along a
trajectory of $5\times 10^6$ iterations reached after a transient time
of $1.2 \times 10^7$ iterations. In Fig. \ref{fig4}(d) it is possible
to note that the time average $\langle p_1 \rangle + \langle p_2
\rangle$ is the same for the two periodic attractors which have period
$2$. However, they can be differentiated by the respective values of
$\lambda_1$. 

Figures~\ref{fig3}(c) and \ref{fig3}(d) show that the duplication of
the region with negative RC is due to the birth of the black region in
which a hyperchaotic and a period-2 attractor coexist, scenario
displayed in Fig.~\ref{fig4}(b) for the case with $c=5\times
10^{-2}$. In this region of the parameter space, the hyperchaotic
attractor [blue color in Fig.~\ref{fig4}(b)] has time average $\langle
p_1 \rangle + \langle p_2 \rangle \approx -2$, and for the period-2
attractor [black color in Fig.~\ref{fig4}(b)] this value is around
$-4\pi$. Since the basin of attraction of the period-2 attractor is
larger ($99\%$ of the $10^4$ ICs used) than the basin of attraction of
the hyperchaotic one, this parametric domain is characterized by a
negative RC. 

\section{Coupling of two different RMs}
\label{diff}
The next step is to keep fixed the values $(K_1,\gamma_1)$ for the map 
$\mbox{\bf RM}^{(1)}$ and change the parameters $(K_2,\gamma_2)$. The
values $(K_1,\gamma_1)$ were chosen according to the value of
$\mathcal{J}_1$. The purpose of this study is to determine how the
current of the particle $i=1$ influences the current $\mathcal{J}_T$
of the whole system. The parametric combinations used for this
analysis are indicated in Fig.~\ref{fig1}(a) by the white points A, B,
C, D, and E, and Table~\ref{tab1} specifies the period of the orbit
obtained and the value of $\mathcal{J}_1$. 
%
\begin{table}[!b]
\caption{\label{tab1} Fixed values used for $K_1$ and $\gamma_1$ in
  model (\ref{rm-el}) and marked as white points in Fig.~\ref{fig1}(a).} 
\label{t1}
\begin{center}
\begin{tabular}{|c|c|c|c|c|}
\hline
Label & $(K_1,\gamma_1)$ & Period & $\mathcal{J}_1$ \\ \hline
A     & $(7.0,0.52)$     & chaos  & $0.02869127$    \\ \hline
B     & $(5.2,0.50)$     & $1$    & $6.28318531$    \\ \hline
C     & $(6.8,0.70)$     & $1$    & $12.56637061$   \\ \hline
D     & $(5.6,0.64)$     & $2$    & $-6.28318531$   \\ \hline
E     & $(6.1,0.81)$     & $2$    & $-12.56637061$  \\ \hline
\end{tabular}
\end{center}
\end{table}

\subsection{$\mbox{\bf RM}^{(1)}$ with $\mathcal{J}_1 \sim 0$}

The first case analyzed is $(K_1,\gamma_1)=(7.0,0.52)$, for which a chaotic 
dynamics is obtained for the map $\mbox{\bf RM}^{(1)}$ and the RC $\mathcal{J}_1$ 
is around zero. The parameter space $(K_2,\gamma_2)$ is displayed in 
Figs.~\ref{fig5}(a), \ref{fig5}(b), and \ref{fig5}(c), with coupling strengths 
$c=10^{-7},10^{-5}$ and $c=10^{-3}$, respectively. In Fig.~\ref{fig5}(a) it is 
possible to note that weak couplings between the particles have a similar effect 
on the ISSs as weak noise, namely to destroy only the antennas of the ISSs. The 
same is observed for the case $(K_1,\gamma_1)=(K_2,\gamma_2)$.  

\begin{widetext}
$\quad$
\begin{figure}[!t]
  \centering
  \includegraphics*[width=0.95\columnwidth]{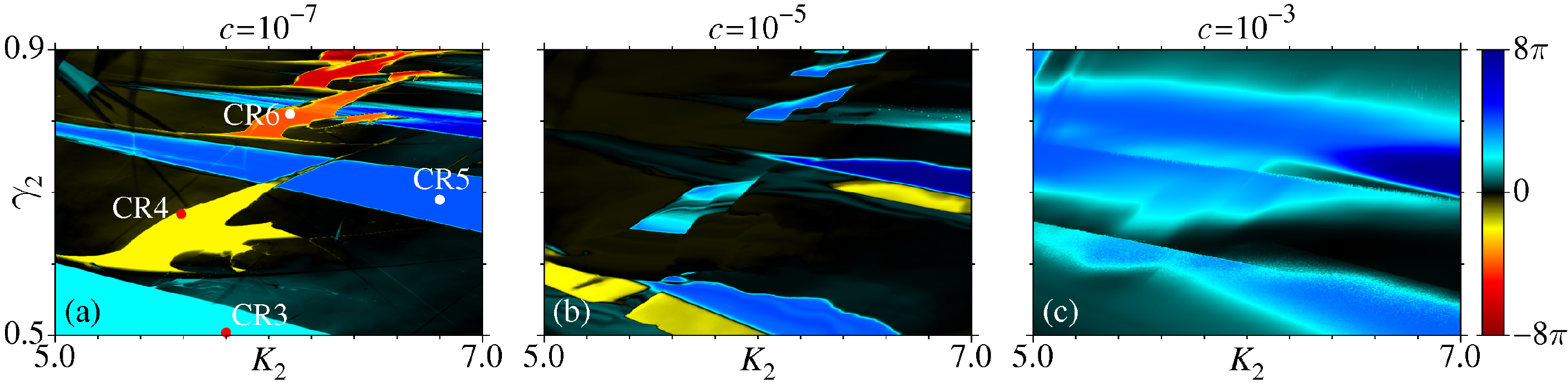}
  \caption{(Color online) Parameter space $(K_2,\gamma_2)$ with colors
    representing the TRC $\mathcal{J}_T$ for the system~(\ref{rm-el}) using 
    (a) $c=10^{-7}$, (b) $c=10^{-5}$, and (c) $c=10^{-3}$. The parametric 
    combination $(K_1,\gamma_1)=(7.0,0.52)$ [white point A in
    Fig.~\ref{fig1}(a)] was kept fixed, giving rise to a chaotic
    dynamics with $\mathcal{J}_1\approx 0$ for the map $\mbox{\bf
      RM}^{(1)}$, according to Table~\ref{t1}.}  
  \label{fig5}
\end{figure}
\end{widetext}

\begin{figure}[!t]
  \centering
  \includegraphics*[width=0.96\columnwidth]{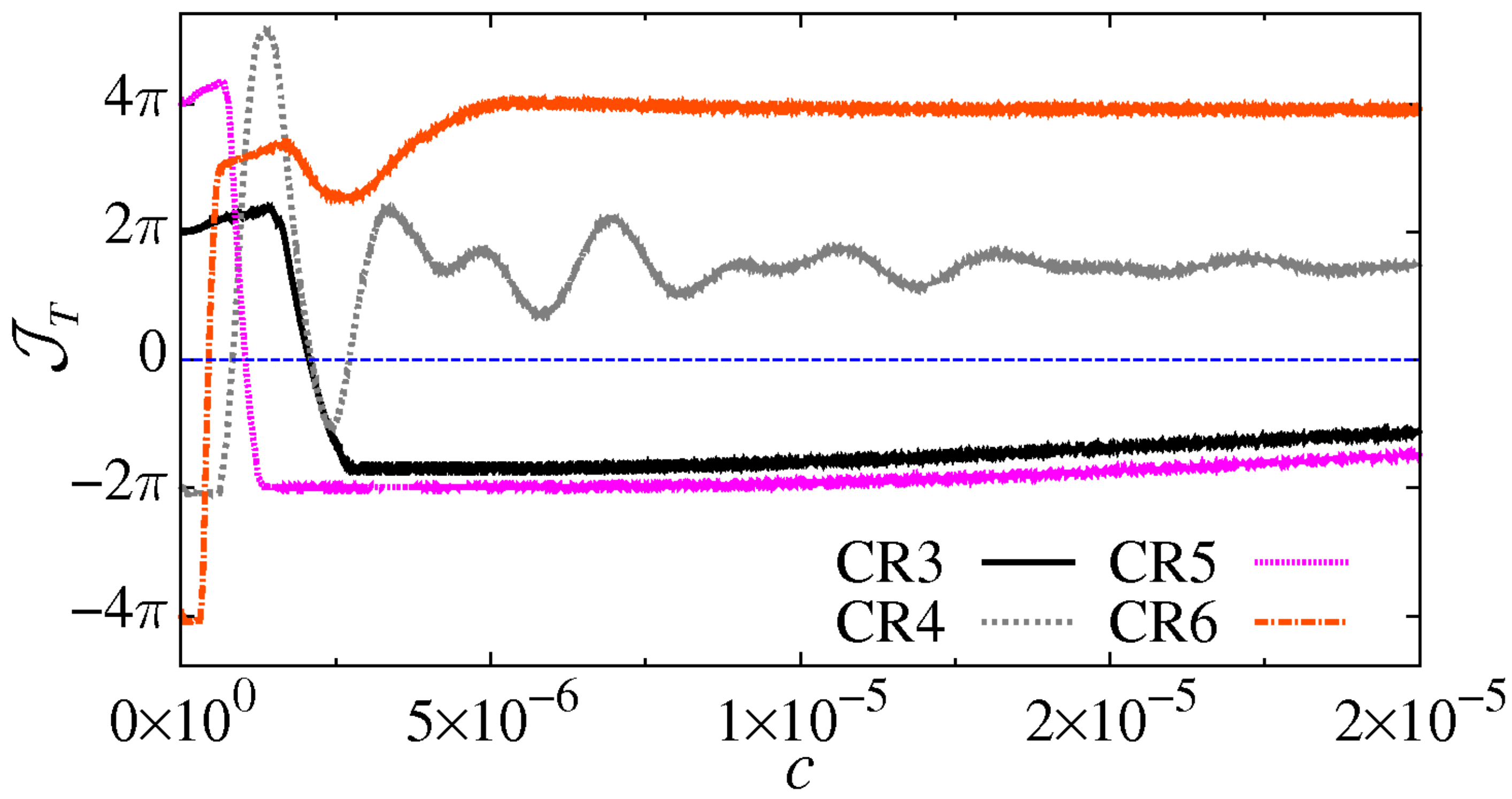}
  \caption{(Color online) TRC $\mathcal{J}_T$ as a function of $c$ for the system 
  (\ref{rm-el}) considering the coupling of a chaotic map with $(K_1,\gamma_1)=
  (7.0,0.52)$ and the map $\mbox{\bf RM}^{(2)}$ with parameters $(K_2,\gamma_2)$ 
  indicated by the white points CR3, CR4, CR5, and CR6 in Fig. \ref{fig5}(a). For 
  each value of $c$, $\mathcal{J}_T$ was obtained using $10^4$ ICs and $N=5\times 
  10^4$ iterations after a transient time of $8\times 10^4$ iterations.}  
  \label{fig6}
\end{figure}

In Fig.~\ref{fig5}(b), for which an intensity $c=10^{-5}$ was applied,
unexpected current reversals can be observed. To treat this phenomenon
in more detail, we indicate in  Fig.~\ref{fig5}(a) four parametric
combinations $(K_2,\gamma_2)$ for which the current reversal
occurs. They  are:
%
(1) CR3: $(K_2,\gamma_2)=(5.8,0.50)$;
(2) CR4: $(K_2,\gamma_2)=(5.6,0.67)$;
(3) CR5: $(K_2,\gamma_2)=(6.8,0.69)$;
and (4) CR6: $(K_2,\gamma_2)=(6.1,0.81)$.
The value of the TRC $\mathcal{J}_T$ as a function of $c$ for these
parametric combinations is plotted in Fig.~\ref{fig6}. This figure
shows that for the points CR3 and CR5 the transition  occurs from
$\mathcal{J}_T>0$ to $\mathcal{J}_T<0$, while for CR4 and CR6  the
transition occurs from $\mathcal{J}_T<0$ to $\mathcal{J}_T>0$. Beside
this, it is possible to note that these transitions happen for weak
couplings ($c\sim 10^{-6}$). Fig.~\ref{fig5}(c) shows that no
regions with negative TRCs can be found anymore when using
$c=10^{-3}$.  
%

\subsection{$\mbox{\bf RM}^{(1)}$ with $\mathcal{J}_1 > 0$}

Figure~\ref{fig7} shows the parameter space $(K_2,\gamma_2)$ of the map
$\mbox{\bf RM}^{(2)}$ coupled to $\mbox{\bf RM}^{(1)}$ with parametric 
combinations $(K_1,\gamma_1)=(5.2,0.5)$ [white point B in Fig.~\ref{fig1}(a)] 
and $(K_1,\gamma_1)=(6.8,0.7)$ [white point C in Fig.~\ref{fig1}(a)], 
both cases leading to a period-1 stable attractor. For the uncoupled
case plotted in Fig.~\ref{fig1}(a), ISSs with period-1 contain  
positive RCs and are organized along a preferential direction. Along this 
direction, the absolute value of the RC increases $2\pi$ inside ISSs of same 
period \cite{Alan-2011}. For instance, the parametric combinations indicated 
by the white points B and C in Fig.~\ref{fig1}(a) have RCs $\mathcal{J}_1
\approx 2\pi$ and $\mathcal{J}_2\approx 4\pi$, respectively, according to Table 
\ref{tab1}. Keeping fixed the parameter-pair
$(K_1,\gamma_1)=(5.2,0.5)$ in model~(\ref{rm-el}), the results are
shown in Figs.~\ref{fig7}(a) for $c=10^{-7}$, \ref{fig7}(b) for
$c=10^{-5}$, and \ref{fig7}(c) for $c=10^{-3}$.

In Fig.~\ref{fig7}(a) we can see that, by keeping fixed the value 
$\mathcal{J}_1  \approx 2\pi$ for the map $\mbox{\bf RM}^{(1)}$, it is 
possible to generate a TRC $\mathcal{J}_T = 0$ (black color) in the 
parametric domain inside the ISS that originally contain RC with value around 
$-2\pi$ in the uncoupled case. Increasing the coupling between the maps to
$c=10^{-5}$ and to $c=10^{-3}$, the ISSs  with negative TRC (yellow
and red regions) collapse to a single region of the parameter
space. For $c=10^{-3}$ [Fig. \ref{fig7}(c)], the positive TRCs
indicated by the different shades
\begin{widetext}
$\quad$
\begin{figure}[!b]
  \centering
  \includegraphics*[width=0.95\columnwidth]{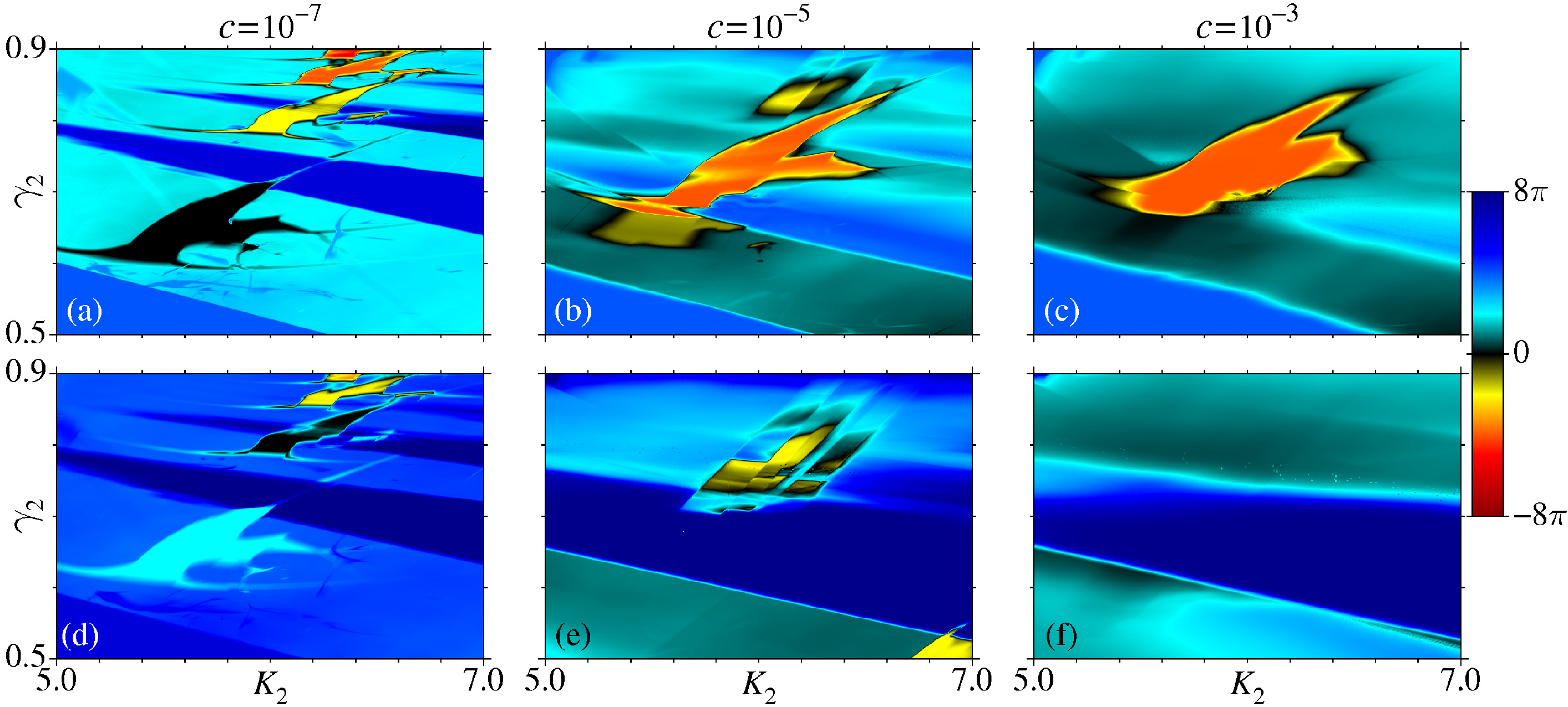}
  \caption{(Color online) Parameter space $(K_2,\gamma_2)$ with colors 
  representing the TRC $\mathcal{J}_T$ for the system
  (\ref{rm-el}). In (a) ($c=10^{-7}$), (b) ($c=10^{-5}$), and (c)
  ($c=10^{-3}$) the parametric combination
  $(K_1,\gamma_1)=(5.2,0.5)$ [white point B in Fig. \ref{fig1}(a)] was
  kept fixed, resulting in orbits with period-1 for the map $\mbox{\bf
    RM}^{(1)}$ and in a RC $\mathcal{J}_1\approx 2\pi$, according to
  Table~\ref{t1}. In (d) ($c=10^{-7}$), (e) ($c=10^{-5}$), and (f)
  ($c=10^{-3}$) the parametric combination $(K_1,\gamma_1)=(6.8,0.7)$
  [white point C in Fig. \ref{fig1}(a)] was kept fixed, resulting in
  orbits with period-1 for the map $\mbox{\bf RM}^{(1)}$ and 
  in a RC $\mathcal{J}_1\approx 4\pi$, according to Table~\ref{t1}.} 
  \label{fig7}
\end{figure}
\end{widetext}
of blue occupy the most part of the parameter space $(K_2,\gamma_2)$.
Choosing $(K_1,\gamma_1)=(6.8,0.7)$, for which $\mathcal{J}_1 \approx 4\pi$, similar 
results are obtained. In Fig. \ref{fig7}(d), which shows the case $c=10^{-7}$, we can 
note a zero TRC $\mathcal{J}_T$ (black color) in the region inside the ISS that has 
RC around $-4\pi$ for the uncoupled case. Using $c=10^{-5}$, case plotted in Fig. 
\ref{fig7}(e), only a small parametric domain lead to negative TRCs, while using 
$c=10^{-3}$, case showed in Fig.~\ref{fig7}(f), only positive TRCs can be found. 

\subsection{RM$^{(1)}$ with $\mathcal{J}_1 < 0$}

Now we present in Fig.~\ref{fig8} the results obtained by keeping fixed $(K_1,
\gamma_1)=(5.6,0.64)$ [white point D in Fig.~\ref{fig1}(a)] and $(K_1,\gamma_1)=
(6.1,0.81)$ [white point E in Fig.~\ref{fig1}(a)] for the map 
$\mbox{\bf RM}^{(1)}$. Considering the uncoupled case ($c=0$), the resulting RCs 
for these parametric combinations are $\mathcal{J}_1\approx -2\pi$ and 
$\mathcal{J}_1\approx -4\pi$, in this order, both cases associated with period-2 
attractors (see Table~\ref{tab1}). Figure~\ref{fig8}(a), obtained using 
$(K_1,\gamma_1)=(5.6,0.64)$ and $c=10^{-7}$, shows that the period-1 ISS which 
contain positive RC around $2\pi$ in the uncoupled case now has a TRC 
$\mathcal{J}_T=0$ (black color), since we kept fixed the RC $\mathcal{J}_1\approx 
-2\pi$. In Figs.~\ref{fig8}(b) and \ref{fig8}(c), which show the cases $c=10^{-5}$ 
and $c=10^{-3}$, respectively, it is possible to note that regions with positive 
TRCs (blue color) remain occupying considerable portions of the parameter space, 
even though we set a negative RC for the map $\mbox{\bf RM}^{(1)}$. This phenomenon 
can be understood when considering that ISSs with lower periods are more resistant 
to perturbative effects, being the period-1 ISSs less affected than the period-2 
ISSs, for example. This explains why in Fig.~\ref{fig7} regions with
negative TRCs (associated to the period-2 ISSs of the uncoupled case)
tend to vanish  whenever a positive value for $\mathcal{J}_1$ was set
(picked  up inside the period-1 ISSs of the uncoupled case), while in
Fig.~\ref{fig8} a  large region with positive TRCs (cyan and blue)
remain for considerable values of coupling strength even though we set
negative values for $\mathcal{J}_1$. 

In Figs.~\ref{fig8}(d), \ref{fig8}(e), and \ref{fig8}(f) the results obtained
keeping fixed the parameters $(K_1,\gamma_1)=(6.1,0.81)$ and using $c=10^{-7}, 
10^{-5}$, and $c=10^{-3}$, respectively, are shown. In this case, a null TRC 
takes place inside the period-1 ISS for which a positive RC of magnitude around 
$4\pi$ is obtained for $c=0$, indicated by the black color. In Fig.~\ref{fig8}(e) 
it is possible to find a large region with high values of negative TRC (dark-red 
color). Increasing the coupling parameter to $c=10^{-3}$ such region disappear, 
and surprisingly positive TRCs are present in a wide portion of the parameter 
space, which is related to the robustness of the period-1 ISSs of the uncoupled 
case.   

\begin{widetext}
$\quad$
\begin{figure}[!t]
  \centering
  \includegraphics*[width=0.95\columnwidth]{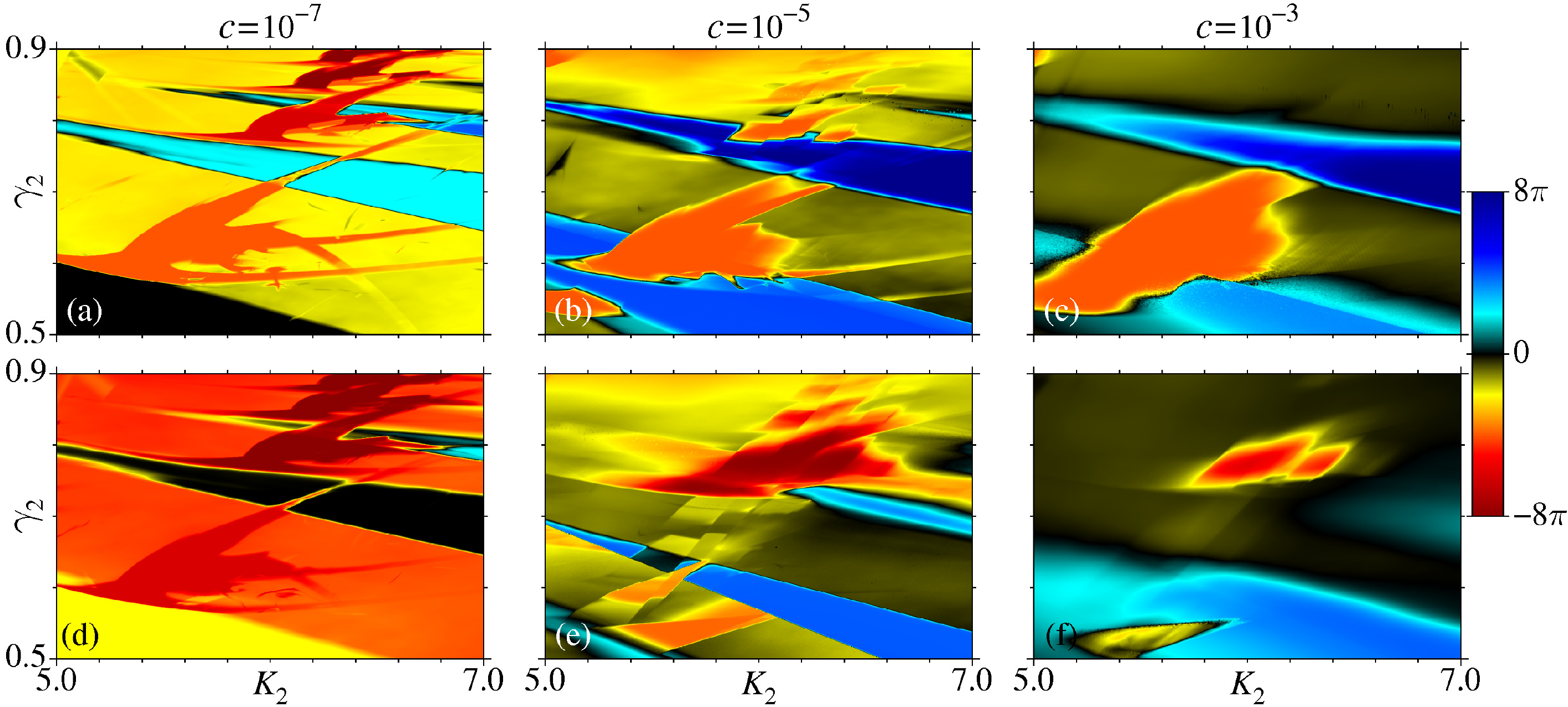}
  \caption{(Color online) Parameter space $(K_2,\gamma_2)$ with colors 
  representing the TRC $\mathcal{J}_T$ for the system (\ref{rm-el}). In (a) 
  ($c=10^{-7}$), (b) ($c=10^{-5}$) and (c) ($c=10^{-3}$) the parametric combination 
  $(K_1,\gamma_1)=(5.6,0.64)$ [white point D in Fig. \ref{fig1}(a)] was kept
  fixed, resulting in orbits with period-2 for the map $\mbox{\bf RM}^{(1)}$ 
  and in a RC $\mathcal{J}_1\approx -2\pi$, according to Table \ref{t1}. In (d) 
  ($c=10^{-7}$), (e) ($c=10^{-5}$) and (f) ($c=10^{-3}$) the parametric combination 
  $(K_1,\gamma_1)=(6.1,0.81)$ [white point E in Fig. \ref{fig1}(a)]
  was kept fixed, resulting in orbits with period-2 for the map
  $\mbox{\bf RM}^{(1)}$ and in a RC $\mathcal{J}_1 \approx -4\pi$,
  according to Table~\ref{t1}.} 
  \label{fig8}
\end{figure}
\end{widetext}

\section{Conclusions}
\label{conclusion}

Recently it was shown \cite{Alan-2011} that single ratchet devices
produce optimal currents when the parameters of the system are chosen
inside ISSs which live in the parameter space and generate a stable dynamics. 
We recall here that the ISSs explain all the relevant dynamics which
occurs to the ratchet current, ranging from optimal current, current
reversal,  temperature induced destruction or enhancement of the
current, among others. The present work analyses the relevance of such
ISSs in the case of two coupled ratchet devices. Furthermore, our
results show that it is possible to use one ratchet device
to control the current of the other one, leading to completely
unexpected behaviors.

We study numerically some of the different dynamical behaviors of a
system  composed of two elastically coupled ratchet. Such a system can
exhibit, depending on the intensity $c$ of coupling, current reversals
(from $\mathcal{J}_T>0$ to $\mathcal{J}_T<0$ and vice versa),
duplication of ISSs, hyperchaos and the coexistence of the different
kind of attractors, namely chaotic and periodic, periodic and
hyperchaotic, chaotic and hyperchaotic, quasiperiodic and periodic, 
and more than one periodic attractor. The variety of complex dynamics 
induced by the interaction between the ratchets can be nicely explained 
in terms of the ISSs. However, when the 
value of coupling $c$ between the maps $\mbox{\bf RM}^{(1)}$ and $\mbox{\bf
RM}^{(2)}$ increases, optimal RCs can appear in distinct regions of the 
parameter space, differently from the case with one uncoupled particle in which
high values of RCs are found only inside the ISSs. Finally, the duplication of 
the ISSs caused by the interacting ratchet (observed using external forces in 
1- and 2-dimensional maps \cite{rafael17-2, rafael17-1, RMS-2018454} and in
continuous-time dynamical systems \cite{PRE98-032210}) is a very
interesting results  and leads to an intriguing question regarding the
dynamics of many particles interacting systems. It is possible to
observe multiplication of ISSs when  multiple ratchet are weakly
coupled? If so, multiple particle coupled ratchet  devices should have
an almost regular behavior with optimal current. This is  
subject of future work.

We also mention that in this work we used equal coupling
strength $c$ between both ratchet. In case of distinct coupling
strengths, say $c_{\rm RM1} \neq c_{\rm RM2}$ (even if one of 
them is very small), might drastically change the configuration 
of the  attractors and multistable behaviors in phase space. Interesting
results about this subject were obtained for two elastically
interacting neurons modeled by FitzHugh-Nagumo oscillators coupled
to each other \cite{Campbell-2001}. There, it was shown that when the
coupling strength between neurons are different, multistability still
holds, but the attractors configuration is completely changed. A
similar scenario is expected in problems involving near identical
ratchet oscillators coupled together with different strengths, but it is
very difficult to present some generic statement about how its dynamics
should be affected. Indeed, further studies on such direction are
needed.
  
Another important issue is the influence of the noise on the
current of coupled ratchet particles. In this scenario, it is
expected that Levy noise may enhance the transport of particles
\cite{Kurths-2017-1,Kurths-2017-2}, while for Gaussian noise the
total ratchet current tends to zero with increasing noise intensity
since the ISSs of the individual maps
decrease \cite{Alan-2013,RMS-2018454}.

\vspace*{1cm}
\acknowledgments{The authors thank CNPq (Brazil) for financial support
  and, they also acknowledge computational support from Prof.~C.M. de
  Carvalho at LFTC-DFis-UFPR (Brazil). C.M. also thanks FAPESC and
  CAPES (Brazilian agencies) for financial support.} 


\end{document}